\pdfoutput=1
\documentclass[twocolumn]{aa}
\usepackage{pslatex}
\usepackage{graphicx}
\usepackage{txfonts}
\usepackage{amssymb}
\usepackage{wasysym}
\usepackage{times}
\usepackage{natbib}
\bibpunct{(}{)}{;}{a}{}{,}
\usepackage{color}
\usepackage{xspace}
\usepackage{ulem}
\usepackage{setspace} 

\def\DS{D_\mathrm{S}}
\def\DL{D_\mathrm{L}}

\def\ThE{\theta_\mathrm{E}}

\def\rhoS{\rho_\mathrm{*}}
\def\tE{{t_{\rm E}}}
\def\to{{t_{\rm 0}}}
\def\uo{{u_{\rm 0}}}

\def\z{{z}}

\newcommand\Eq[1]{Eq.~(\ref{#1})}
\newcommand\Fig[1]{Fig.~\ref{#1}}

\newcommand\Sec[1]{sec.~\ref{#1}}
\def\ie{{\it i.e.}\xspace}
\def\cf{{\it cf.}\xspace}
\def\d{\mathrm{d}} 

\def\inn{{\rm in}}
\def\out{{\rm out}}

\usepackage{ulem}
\usepackage{color}

\begin{document}
   
\title{Bayesian analysis of caustic-crossing microlensing events}  
\titlerunning{Bayesian analysis of caustic-crossing microlensing events}
\authorrunning{A.~Cassan et al.}
\author{A.~Cassan\inst{1,2}, K.~Horne\inst{3}, N.~Kains\inst{3}, Y.~Tsapras\inst{4,5}, P.~Browne\inst{3}}
\institute{
  {Institut d'Astrophysique de Paris, UMR 7095 CNRS, 98 bis boulevard Arago, 75014 Paris, France} \and
  {Universit\'e Pierre \& Marie Curie, Paris, France} \and
  {Scottish Universities Physics Alliance, School of Physics \&
    Astronomy, University of St Andrews, North Haugh, KY169SS, UK } \and
  {Las Cumbres Observatory, 6740B Cortona Dr, suite 102, Goleta, CA
    93117, USA} \and
  {School of Mathematical Sciences, Queen Mary University of London,
    Mile End Road, London E1 4NS, UK}
} 
\date{Received ; accepted}
 \abstract
 {}
 {
   Caustic-crossing binary-lens microlensing events are important
   anomalous events because they are capable of detecting an
   extrasolar planet companion orbiting the lens star.
   Fast and robust modelling methods are thus of prime interest 
   in helping to decide whether a planet is detected by an event. 
   Cassan introduced a new set of parameters to model
   binary-lens events, which are closely related to properties
   of the light curve. In this work, we explain how Bayesian priors can be
   added to this framework, and investigate on interesting 
   options.
 } 
 {
   We develop a mathematical formulation that allows us to compute
   analytically the priors on the new parameters, given some previous
   knowledge about other physical quantities. We explicitly compute the priors
   for a number of interesting cases, and show how this can be
   implemented in a fully Bayesian, Markov chain Monte Carlo algorithm.
 }
 {
   Using Bayesian priors can accelerate microlens fitting codes by
   reducing the time spent considering physically implausible models,
   and helps us to discriminate between alternative models based on
   the physical plausibility of their parameters.
 } 
 {}
 \maketitle

 \section{Introduction} \label{sec:caustics}
 
 \citet{MaoPaczynski1991} first suggested that observations of Galactic 
 gravitational microlensing events could lead to the discovery of extrasolar
 planets. Microlensing involves the time-dependent brightening and then
 dimming of a background source star as an intervening massive
 object (the lens) crosses the observer line-of-sight. Light rays from
 the source bend in the vicinity of the lens, 
 focusing them toward the observer. 
 Since 1994, survey teams such as
 OGLE\footnote{http://www.astrouw.edu.pl/$\sim$ogle}
 \citep[OGLE~III,][]{Ref-OGLE} 
 and MOA\footnote{http://www.phys.canterbury.ac.nz/moa} 
 \citep{Ref-MOA} have reported more than four thousand microlensing 
 events toward the Galactic bulge to date. Several hundreds of these events 
 have been carefully selected and densely sampled by follow-up networks such 
 as PLANET\footnote{http://planet.iap.fr}, 
 $\mu$FUN\footnote{http://www.astronomy.ohio-state.edu/$\sim$microfun},
 RoboNet\footnote{http://robonet.lcogt.net}, and 
 MiNDSTEp\footnote{http://www.mindstep-science.org}. 
 Although microlensing teams have 
 so far published only nine exoplanet detections, the method itself stands out
 because of its high sensitivity to low-mass planets with orbits 
 of several astronomical units. It thus probes
 in the planet mass-separation plane a region beyond reach of any
 other technique, as demonstrated by the 
 detection of the very first cool super-Earth, OGLE~2005-BLG-390Lb
 \citep{OGLE05390Lb,Jovi2008}.
 
 A number of microlensing events exhibit anomalous behaviour \citep[\ie, they 
 cannot be adequately modelled by the standard single-lens light curve,
 e.g.,][]{Paczynski1986} and some of these anomalies can be attributed to
 lensing by binary objects. The types of light curves produced by
 binary lensing form a rich tapestry but, in general, binary systems
 with two equal mass components tend to exhibit pronounced,
 long anomalies in their light curves, whereas when the secondary companion is only a 
 small fraction of the total mass, the anomalies can be quite short and subtle.
 It is primarily these latter types of anomalies that may be caused by star-planet 
 binaries \citep{MaoPaczynski1991,GouldLoeb1992}. 
 Nevertheless, because the true nature of the anomaly
 cannot always be established while the microlensing event is still
 ongoing, every binary-lens microlensing event constitutes a prime target for
 planet hunting.

 In binary lensing, the lens system configuration delineates
 regions of space on the source plane that are bound by gravitational
 caustics. Caustics are closed curves with concave segments that meet in
 outward pointing cusps, defined by the location where the Jacobian
 determinant of the lens mapping equation vanishes, \ie, are lines of
 infinite point-source magnification.
 There are three kinds of caustic topologies, which depend on the values of
 the binary lens mass ratio $q$ and the two component projected
 separation $d$ in angular Einstein ring radius $\ThE$
 \citep{Einstein1936}
 \begin{equation}
   \ThE = \sqrt{\frac{4GM}{c^2}
     \left(\frac{\DS-\DL}{\DS\,\DL}\right)}\, , 
 \end{equation}
 where $\DS$, $\DL$ are the observer-source and observer-lens distances
 and $M$ the lens total mass. In the close separation regime
 \citep[\cf Fig.~1 of][]{Causfix}, there are three caustics, one
 central (4-cusp) and two (3-cusp) planetary caustics. In the
 intermediate regime, there is only one (6-cusp) caustic, and in the
 wide separation regime, there is one central and one planetary caustic
 (both with 4 cusps).
 
 In many cases, the source trajectory happens to cross a caustic. As
 the source crosses the caustic curve and enters the enclosed area, a
 new pair of images appears, causing a sudden increase in the observed
 brightness. In a similar way, when the source exits the area defined by the
 caustics, the two images merge and disappear, causing a rapid drop in
 the observed brightness. These dramatic changes in
 magnification result in readily recognisable jumps in microlensing
 light curves. As emphasised by \citet{Causfix}, the ingress and
 egress times $t_\inn$ and $t_\out$ may be restricted to within very tight
 intervals when caustic crossing features have been identified in the
 light curve, and thus advantageously used as alternative modelling
 parameters. 

 The new set of binary-lens modelling parameters introduced by
 \citet{Causfix} have the advantage that two of these 
 parameters are very closely related
 to features that can be directly identified in the light
 curve. Using this new formulation to analyse the data of OGLE~2007-BLG-472 in 
 its most straightforward implementation as a maximum likelihood analysis 
 (``minimising $\chi^2$''), \citet{Kains2009} unveiled a subtle
 aspect of binary-lens modelling: relatively improbable physical
 models with very large values of $\tE$ were found with $\chi^2$
 values lower than other more plausible models.  
 To avoid finding parameter combinations that are physically
 unlikely, dramatic progress can be achieved by switching to a Bayesian
 analysis. This is desirable as the Bayesian approach makes use of prior 
 information on the underlying physical parameters, while $\chi^2$ says 
 nothing about parameter plausibility. 

 In this article, we show how to derive Bayesian priors for the
 caustic-crossing binary-lens parameters defined by \citet{Causfix}. These are 
 based on physical priors on quantities that can be estimated from Galactic 
 models or calculated from already observed events (\Sec{sec:MLvsBay} and
 \ref{sec:Bayesprior}). In \Sec{sec:MCMC}, we describe an
 implementation of this Bayesian formalism within a Markov chain
 Monte Carlo fitting scheme, using in particular priors on the
 Einstein time $\tE$ (time for the source to travel an angular
 distance $\ThE$).

 \section{Maximum likelihood  \textit{versus} Bayesian fitting} \label{sec:MLvsBay}
 
 \cite{Causfix} introduced a new parameterisation of the binary
 lens microlens light curve model that is well suited to
 describing caustic-crossing events. In this formalism, the caustic curve in the
 source plane is parameterised by a curvilinear abscissa (or arc
 length) from $0$ to $2$. The trajectory of a source
 crossing a caustic, which is classically parameterised by its
 impact parameter $\uo$ and position angle $\alpha$, can alternatively
 be defined by giving the values $s_\inn$ at ingress and $s_\out$ at 
 egress\footnote{We use the notations ``in'' and ``out'' in place of
   ``entry'' and ``exit'' of \cite{Causfix} to write more condensed
   formulae.}. 
 The two parameters timing the trajectory, $\tE$ (time to cross
 one Einstein radius) and $\to$ (date at minimum impact parameter
 $\uo$), are then replaced by the ingress and 
 egress times $t_\inn$ and $t_\out$. The caustic curve is specified in
 the source (\ie, caustic) plane by a complex function $\zeta(s)=\xi(s)+i\eta(s)$
 (see \Sec{sec:analytic}), and once $s_\inn$ and $s_\out$ are specified,
 the source trajectory is fully defined.
 This bijective switch of parameters, 
 $(\uo,\alpha,\tE,\to) \mapsto (s_\inn,s_\out,t_\inn,t_\out)$, 
 takes advantage of the relatively high precision with which
 $t_\inn$ and $t_\out$ can be inferred from the observations 
 \citep{Kains2009,Kubas2005}.

 Using these new parameters, \cite{Kains2009} analysed the caustic
 crossing event OGLE~2007-BLG-472. The approach taken was
 a maximum likelihood procedure, quantifying the ``goodness-of-fit''
 by a $\chi^2$ statistic, and
 minimising the $\chi^2$ to optimise the fit.
 A grid search in $(d,q)$ with even spacing in $\log{d}$ and $\log{q}$
 was conducted. For each $(d,q)$ caustic configuration, a genetic
 algorithm was used to explore widely the remaining parameter space. 
 While $(s_\inn,s_\out)$ covered the full range of possibilities,
 $[0,2]\times[0,2]$, $t_\inn$ and $t_\out$ evolved in very tight
 intervals based on the values inferred from the light curve features
 (caustic crossing magnification peaks). 
 These first fits were refined using a Markov chain Monte Carlo (MCMC) 
 algorithm, again holding $(d,q)$ fixed while optimising the remaining
 parameters. The best-fit models in each of the identified best-fit
 regions were then found by allowing all parameters to vary.
 
 As expected for binary lens events, the resulting $\chi^2(d,q)$ maps
 uncovered a variety of widely-separated model parameter regions 
 where a relatively low $\chi^2$ could be achieved.
 The lowest $\chi^2$ models corresponded to very low $q$,
 in the planet-mass regime. But with a short duration between the
 caustic entry and exit, and a planetary caustic size scaling as
 $q^{1/2}$, these models implied an extremely long Einstein time 
 $\tE \sim (t_\out-t_\inn)/q^{1/2} \sim 10^4$~days, which
 is very unlikely according to kinematics of stars motions within the
 Milky Way. These best-fit maximum likelihood models were therefore
 rejected on this physical argument.
 This need to reject the lowest $\chi^2$ models
 highlights a weakness in the maximum likelihood approach,
 which neglects prior distributions on the parameter space.
 On the other hand, Bayesian parameter estimation takes proper
 account of prior distributions in the parameter space 
 \citep[see e.g.,][for a review of astrophysical applications]{TrottaMCMC2008}.
 
 In a Bayesian analysis, the posterior probability distribution over 
 the model parameters $\theta$ is a function of the data $D$
 \begin{equation}
   P(\theta|D) = \frac{P(D|\theta) P(\theta)}
   {\int P(D|\theta) P(\theta)\,\d\theta} \,,
 \end{equation}
 where $P(\theta)$ is the prior probability distribution on the 
 parameters, and the denominator partition function
 ensures proper normalisation of the posterior as a probability distribution
 over the parameters $\theta$.
 The likelihood $P(D|\theta)$ is a function of the parameters $\theta$
 and a probability distribution over the data $D$.
 For Gaussian measurement errors with $N$ data points having
 individual standard deviations $\sigma_i$, the likelihood is
 \begin{equation}
   L(\theta) \equiv P(D|\theta) =
   \frac{ \exp{ \left\{ - \frac{1}{2}\chi^2\right\} } }
   {\left(2\pi\right)^{N/2}\prod_{i=1}^{N}\sigma_i} \,.
 \end{equation}
 Since maximising the likelihood corresponds to minimising
 \begin{equation}
   -2\ln{L(\theta)} = \chi^2 + 2 \sum_{i=1}^{N} \ln{ \sigma_i }
   + \frac{N}{2} \ln{ 2\pi } \,,
 \end{equation}
 a maximum likelihood is equivalent to a minimum in $\chi^2$ when the
 error bars $\sigma_i$ are known, and is essentially a Bayesian
 analysis that implicitly assumes a prior that is uniform on the
 chosen parameters intervals. 
 As we show in the next section, assuming more realistic priors
 could substantially affect the fitting process.

 \section{A Bayesian prior for $(s_\inn,s_\out)$} \label{sec:Bayesprior}

 \subsection{Distribution of $(s_\inn,s_\out)$ for isotropic trajectories} \label{sec:trajgame}

 A uniform prior probability distribution in the parameter square
 $(s_\inn,s_\out)$ is implicit in the maximum likelihood
 analysis. Because of the non-linear correspondence between the two
 sets of parameters, it should correspond to a rather unlikely prior
 for the $(\uo,\alpha)$ source trajectory parameters.
 A more plausible prior would for example arrange for the source
 trajectories to be uniformly distributed and isotropic in
 orientation.
  
\begin{figure}[!htbp]
   \begin{center}
     \hspace{0.75cm}\includegraphics[width=7.4cm]{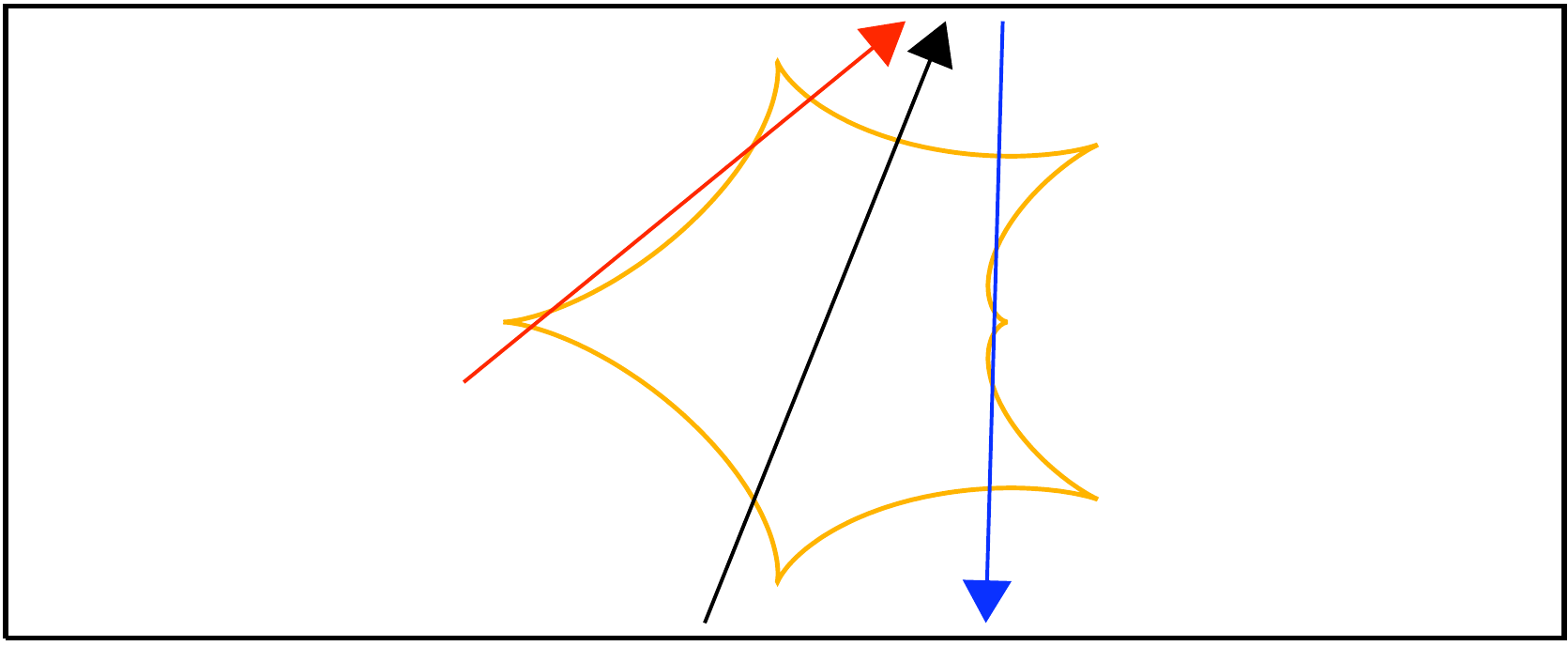} \\
     \includegraphics[width=9cm]{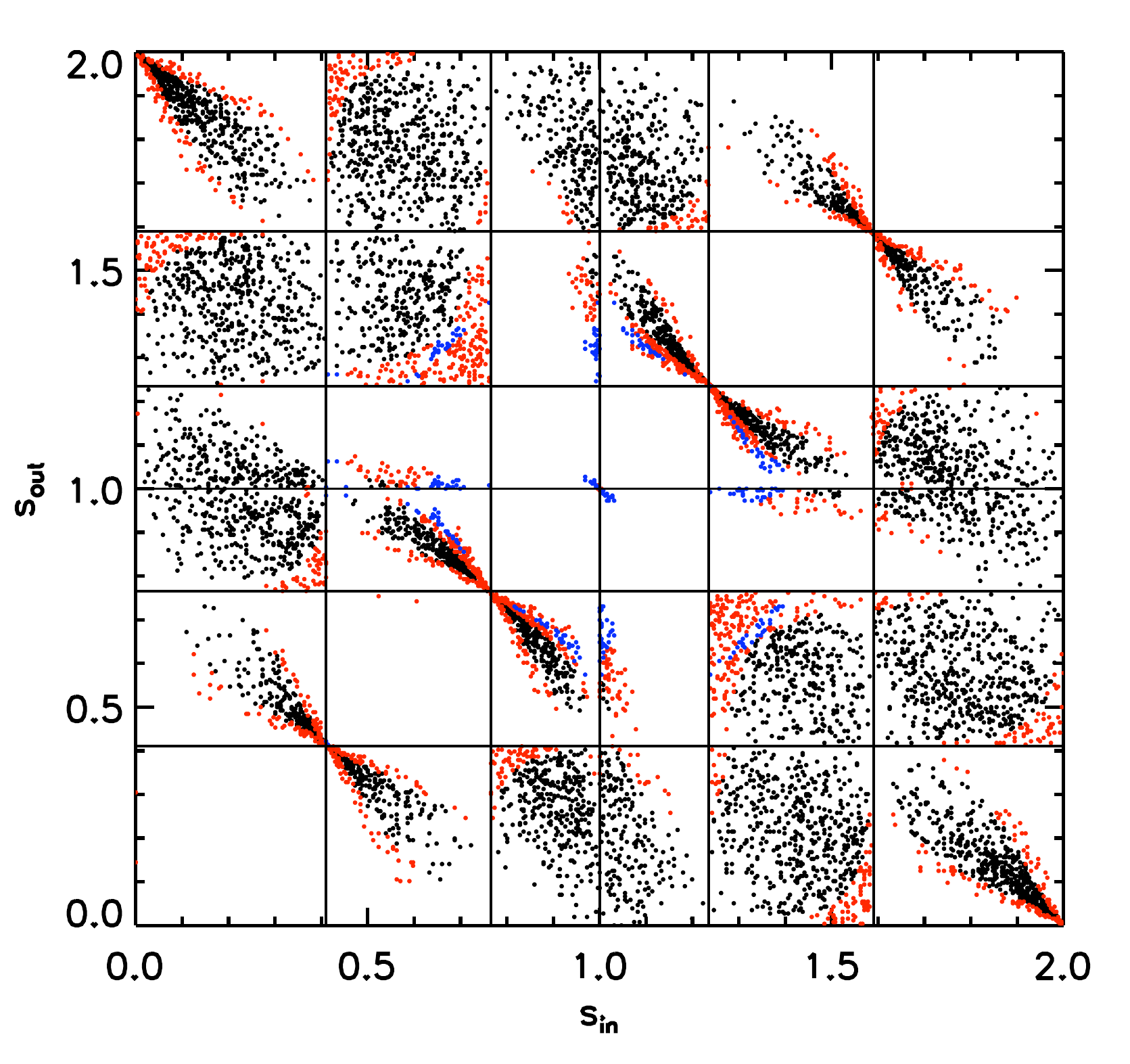}
     \caption{The top panel illustrates
        the three kinds of possible source trajectories crossing a
        caustic: the black line has a single pair of ingress and
        egress points, while the red and blue lines have, respectively,
        two and three ingress/egress points. The bottom panel shows
        in the $(s_\inn,s_\out)$ square the locations of a random
        distribution of $\sim 10^4$ of these trajectories crossing a
        $(d=1.1, q=0,1)$-caustic, with the same colour convention. The
        solid vertical and horizontal black line mark the
        $s$-locations of the caustic cusps.}
     \label{fig:trajshoot}
   \end{center}
 \end{figure}

 In \Fig{fig:trajshoot}, the top panel shows an intermediate
 caustic with $d=1.1$ and $q=0,1$
 (\ie, six cusps, in orange) with several crossing
 trajectories. It can be seen that a straight
 line may cross the caustic at two (black line), four (red line), or
 six (blue line) locations, depending on the number and orientation
 of the cusps. In the bottom panel, $\sim 10^4$ of these trajectories
 were randomly shot and their corresponding position in the
 $(s_\inn,s_\out)$ square reported, using the same colour convention.
 Trajectories with a single pair of ingress and egress map into
 unique black points, while for red and blue trajectory lines, there
 are respectively two and three possible pairs of ingress and egress
 points. 

 We can understand some of the structures in the bottom panel of
 \Fig{fig:trajshoot} as follows:
 vertical and horizontal lines marking the $s$ values at the cusps
 divide the $(s_\inn,s_\out)$ square into boxes. No trajectories appear
 in the boxes along the diagonal because the caustics curve
 concavely outward. It is thus impossible for a line that enters
 at some position between two cusps to exit at any point between 
 those two cusps. In a similar way, other empty regions correspond to
 ingress/egress pairs that cannot be realised by straight lines
 crossing the caustic. 
 The trajectories are seen to bunch up
 around ingress/egress pairs occurring close to a cusp.
 This happens because any trajectory entering close to a cusp is
 very likely (for a wide range of angles) to also exit near
 the same cusp.

 \subsection{Analytical formulation} \label{sec:analytic}
 
 We develop a mathematical formulation that allows us
 to compute analytically priors on $(s_\inn,s_\out)$. The lens
 equation for a binary lens with separation $d$ and mass ratio $q$
 defines the mapping of the position of a point-source $\zeta$ on the
 source plane to the positions of its three or five images at $z$ on
 the lens plane
 \begin{equation}
   \zeta = z - \frac{1}{1+q}\,\left(\frac{1}{\overline{z}}
     + \frac{q}{\overline{z}+d}\right)\,,
 \label{eq:eqlensprim}
 \end{equation}
 where the more massive body is at the centre and the companion on the
 left-hand side. Following \cite{Witt1990}, the caustic lines $\zeta$ are
 parametrised by a parameter $\phi \in [0,2\pi]$
 \begin{equation}
   \frac{1}{1+q}\left[\frac{1}{z^2} +
     \frac{q}{(z+d)^2}\right] = e^{-i\phi} \,,
   \label{eq:Wittprim}
 \end{equation}
 where $z$ and $\zeta$ satisfy \Eq{eq:eqlensprim}. For a given angle
 $\phi$, it is possible to solve a fourth order polynomial equation in $z$
 to obtain the corresponding caustic points $\zeta$. While the
 parameter $\phi$ is used here to write the useful formulae, in
 practice we use instead the equivalent parameter $s = s(\phi)$
 (bijection) introduced by \cite{Causfix}, which has the advantage of
 sampling the caustics evenly. 

 To write more condensed formulae, we use notations that
 resemble two-dimensional vector operations.  
 Given two complex numbers $\zeta_1 = \xi_1 +i \eta_1$ and 
 $\zeta_2 = \xi_2 +i \eta_2$, we write 
 $\zeta_1\wedge\zeta_2 = \xi_1 \eta_2 - \eta_1 \xi_2$ (``wedge
 product'') and 
 $\zeta_1\cdot\zeta_2 = \xi_1 \xi_2 + \eta_1 \eta_2$ 
 (``scalar product''), which are both real numbers. 
 Moreover, a quantity related to a caustic entry (exit)
 is indicated by a subscript ``$\inn$'' (``$\out$'').
 Using the usual convention that $\uo > 0$ when the origin of the
 coordinate system stays on the right-hand side of the source
 trajectory, one can write
 \begin{equation}
   \uo = \frac{\zeta_\out\wedge\zeta_\inn}{|\zeta_\out-\zeta_\inn|} \,,
   \label{eq:uo}
 \end{equation}
 \begin{equation}
   \alpha =
   \arctan\left(\frac{\eta_\out-\eta_\inn}{\xi_\out-\xi_\inn}\right) +
   \pi \, H\left(\xi_\inn-\xi_\out\right) \,,
   \label{eq:alpha}
 \end{equation}
 \begin{equation}
   \tE = \frac{t_\out-t_\inn}{|\zeta_\out-\zeta_\inn|} \,,
   \label{eq:tE}
 \end{equation}
 \begin{equation}
  \to = \frac{t_\out+t_\inn}{2} - (t_\out-t_\inn)
  \left[\frac{1}{2}\frac{\zeta_\out+\zeta_\inn}{\zeta_\out-\zeta_\inn} + 
    i\frac{\zeta_\out\wedge\zeta_\inn}
    {\left|\zeta_\out-\zeta_\inn\right|^2}\right] \,, 
   \label{eq:to}
 \end{equation}
 where $H$ is the Heaviside step function, and 
 $\alpha = \frac{\pi}{2} \, {\rm sign}\left(\eta_\out-\eta_\inn\right)$
 if $\xi_\out=\xi_\inn$.
 The transformation between the two sets of parameters is given by the
 Jacobian
 \begin{equation}
   J = 
   \left|
     \frac{\partial\left(\uo,\alpha,\tE,\to\right)}
     {\partial\left(s_\inn,s_\out,t_\inn,t_\out\right)}
   \right| \,.
 \end{equation}
 Since the dependencies of the classical parameters with respect to the new
 ones are $\uo(s_\inn,s_\out)$, $\alpha(s_\inn,s_\out)$,
 $\tE(s_\inn,s_\out,t_\inn,t_\out)$, and
 $\to(s_\inn,s_\out,t_\inn,t_\out)$, $J$ reads (using $\phi$
 instead of $s$)
 \begin{equation} 
   J = \left|  
     \begin{array}{cccc}
       \frac{\partial\uo}{\partial\phi_\inn} &
       \frac{\partial\uo}{\partial\phi_\out} &
       0 & 0 \\
       \frac{\partial\alpha}{\partial\phi_\inn} &
       \frac{\partial\alpha}{\partial\phi_\out} & 
       0 & 0 \\ 
       \frac{\partial\tE}{\partial\phi_\inn} &
       \frac{\partial\tE}{\partial\phi_\out} &
       \frac{\partial\tE}{\partial t_\inn} &
       \frac{\partial\tE}{\partial t_\out} \\
       \frac{\partial\to}{\partial\phi_\inn} &
       \frac{\partial\to}{\partial\phi_\out} &
       \frac{\partial\to}{\partial t_\inn} &
       \frac{\partial\to}{\partial t_\out} \\
     \end{array} \right| = 
   \left|\frac{\partial\left(\uo,\alpha\right)}
     {\partial\left(\phi_\inn,\phi_\out\right)}\right| \times
   \left|\frac{\partial\left(\tE,\to\right)}
     {\partial\left(t_\inn,t_\out\right)}\right| \,.
 \end{equation}
 After some algebra, we find for the components of the two latter
 Jacobians
 \begin{eqnarray}
   \frac{\partial\uo}{\partial\phi_\inn} &=&
   \frac{\partial\uo}{\partial\xi_\inn}\,\frac{d\xi_\inn}{d\phi_\inn}
   +  
   \frac{\partial\uo}{\partial\eta_\inn}\,\frac{d\eta_\inn}{d\phi_\inn}
   \nonumber \\
   &=&
   \frac{\left(\zeta_\out-\zeta_\inn\right)\cdot\zeta_\out}{\left|\zeta_\out-\zeta_\inn\right|^3} 
   \, \left[\left(\zeta_\out-\zeta_\inn\right) \wedge 
     \frac{d\zeta_\inn}{d\phi_\inn}\right] \,, \\
   \frac{\partial\uo}{\partial\phi_\out} &=&
   - \frac{\left(\zeta_\out-\zeta_\inn\right)\cdot\zeta_\inn}{\left|\zeta_\out-\zeta_\inn\right|^3} 
   \, \left[\left(\zeta_\out-\zeta_\inn\right) \wedge 
     \frac{d\zeta_\out}{d\phi_\out}\right] \,, \\
   \frac{\partial\alpha}{\partial\phi_\inn} &=&
   \frac{\partial\alpha}{\partial\xi_\inn}\,\frac{d\xi_\inn}{d\phi_\inn}
   +  
   \frac{\partial\alpha}{\partial\eta_\inn}\,\frac{d\eta_\inn}{d\phi_\inn}
   \nonumber \\
   &=&
   - \frac{\left(\zeta_\out-\zeta_\inn\right) \wedge
       \frac{d\zeta_\inn}{d\phi_\inn}}{\left|\zeta_\out-\zeta_\inn\right|^2}
   \,, \\
   \frac{\partial\alpha}{\partial\phi_\out} &=&
   \frac{\left(\zeta_\out-\zeta_\inn\right) \wedge
       \frac{d\zeta_\out}{d\phi_\out}}{\left|\zeta_\out-\zeta_\inn\right|^2}
   \,, \\
   \frac{\partial\tE}{\partial t_\inn} &=&
   -\frac{1}{\left|\zeta_\out-\zeta_\inn\right|} \,, \\
   \frac{\partial\tE}{\partial t_\out} &=&
   \frac{1}{\left|\zeta_\out-\zeta_\inn\right|} \,, \\
   \frac{\partial\to}{\partial t_\inn} &=&
   \frac{1}{2} + \left[\frac{1}{2}\frac{\zeta_\out+\zeta_\inn}{\zeta_\out-\zeta_\inn} + 
     i\frac{\zeta_\out\wedge\zeta_\inn}
     {\left|\zeta_\out-\zeta_\inn\right|^2}\right] \,, \\
   \frac{\partial\to}{\partial t_\out} &=&
   \frac{1}{2} - \left[\frac{1}{2}\frac{\zeta_\out+\zeta_\inn}{\zeta_\out-\zeta_\inn} + 
     i\frac{\zeta_\out\wedge\zeta_\inn}
     {\left|\zeta_\out-\zeta_\inn\right|^2}\right] \,, 
 \end{eqnarray}
 so that
 \begin{eqnarray}
   \label{eq:J} 
   \left|\frac{\partial\left(\uo,\alpha\right)}
     {\partial\left(\phi_\inn,\phi_\out\right)}\right|  &=&
   \frac{\left|\left(\zeta_\out-\zeta_\inn\right) \wedge
       \frac{d\zeta_\inn}{d\phi_\inn}\right|
     \left|\left(\zeta_\out-\zeta_\inn\right) \wedge
       \frac{d\zeta_\out}{d\phi_\out}\right|}
   {\left|\zeta_\out-\zeta_\inn\right|^3} \,, \\
   \left|\frac{\partial\left(\tE,\to\right)}
     {\partial\left(t_\inn,t_\out\right)}\right| &=&
   \frac{\frac{\partial\to}{\partial t_\inn} +
     \frac{\partial\to}{\partial t_\out}}
   {\left|\zeta_\out-\zeta_\inn\right|} =
   \frac{1}{\left|\zeta_\out-\zeta_\inn\right|} \,,
 \end{eqnarray}
 which gives
 \begin{equation}
   J =  
   \frac{\left|\left(\zeta_\out-\zeta_\inn\right) \wedge
       \frac{d\zeta_\inn}{d\phi_\inn}\right|
     \left|\left(\zeta_\out-\zeta_\inn\right) \wedge
       \frac{d\zeta_\out}{d\phi_\out}\right|}
   {\left|\zeta_\out-\zeta_\inn\right|^4} \,. 
   \label{eq:Jacobfinal}
 \end{equation}
 The derivatives $d\zeta/d\phi$ evaluated at the caustic entry and
 exit are given by \citep{Causfix}
 \begin{equation}
   \frac{d\zeta}{d\phi} = \frac{d z}{d\phi} +
   e^{i\phi}\,\overline\frac{d z}{d\phi} \,,
   \label{eq:dydphi}
 \end{equation}
 where
 \begin{equation}
   \frac{d\z}{d\phi} = \frac{i}{2}\,\frac{(\z+d)^2+q\,\z^2}
	{(\z+d)^3+q\,\z^3}\,(\z+d)\,\z \,.
 \end{equation}
 In the limit of cusp-crossing trajectories, \ie,
 $\zeta_\out-\zeta_\inn\rightarrow0$, $J$ behaves like
 $1/|\zeta_\out-\zeta_\inn|^2$.

 \begin{figure*}[!htbp]
   \begin{center}
     \includegraphics[width=3.5cm]{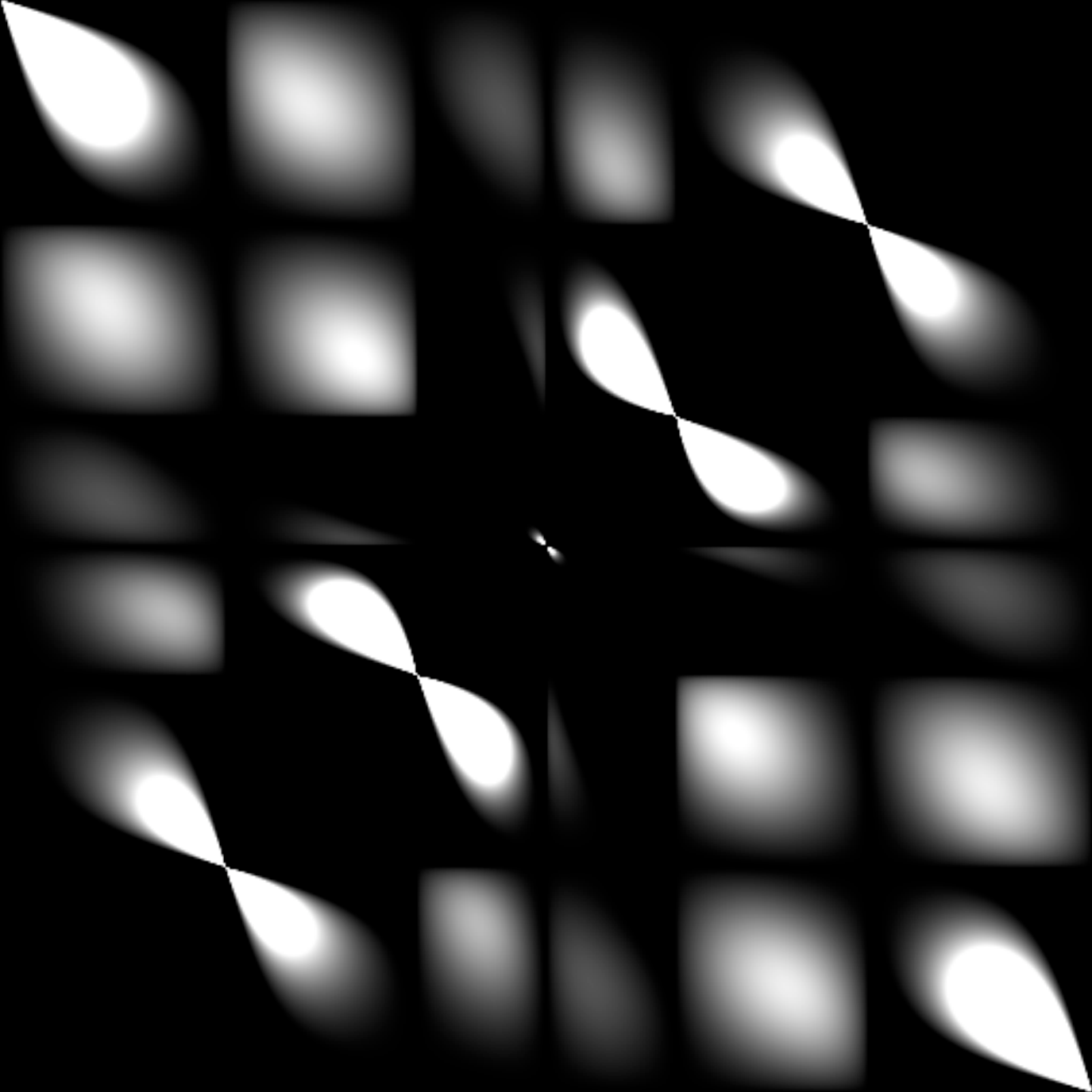} 
     \includegraphics[width=3.5cm]{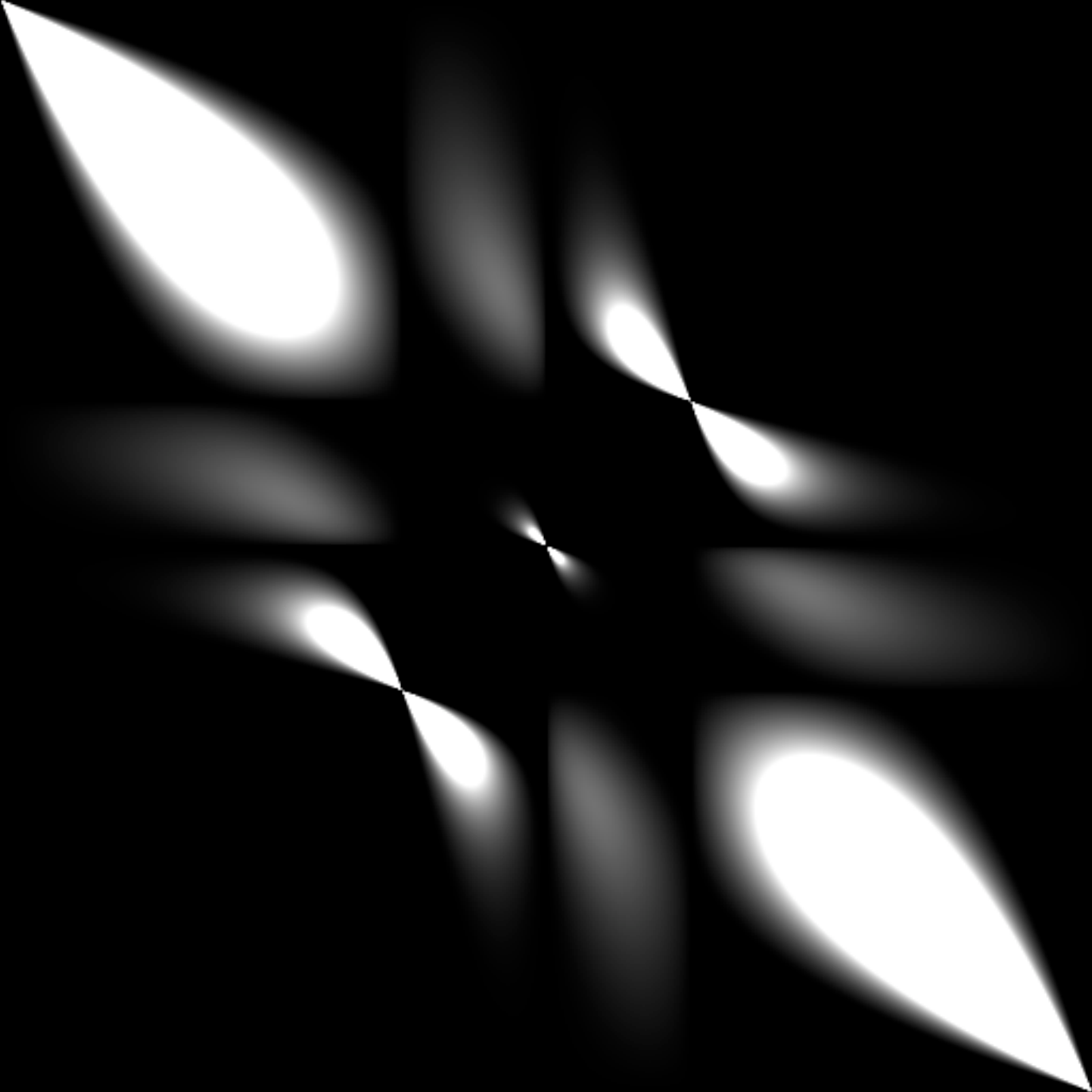}
     \includegraphics[width=3.5cm]{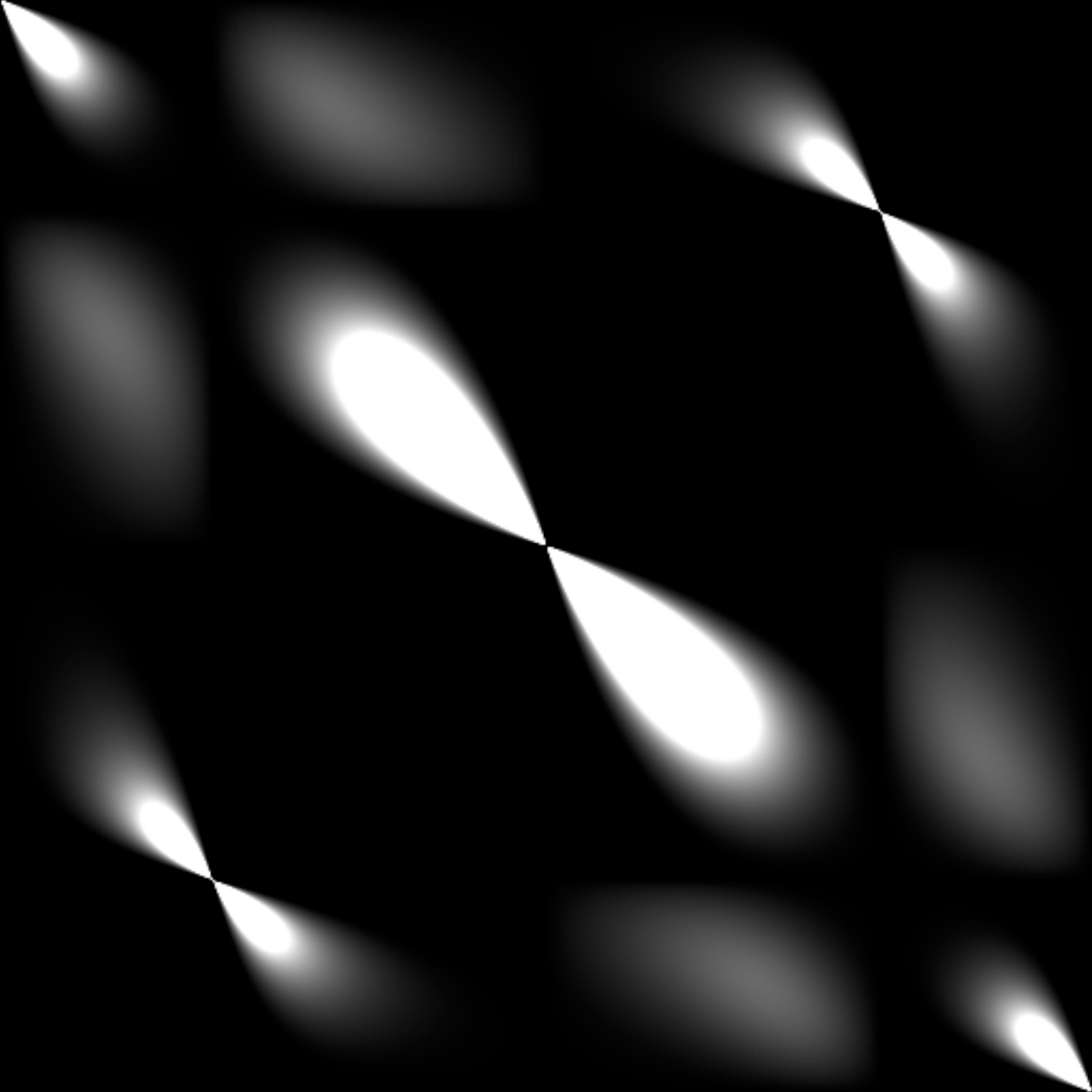}
     \includegraphics[width=3.5cm]{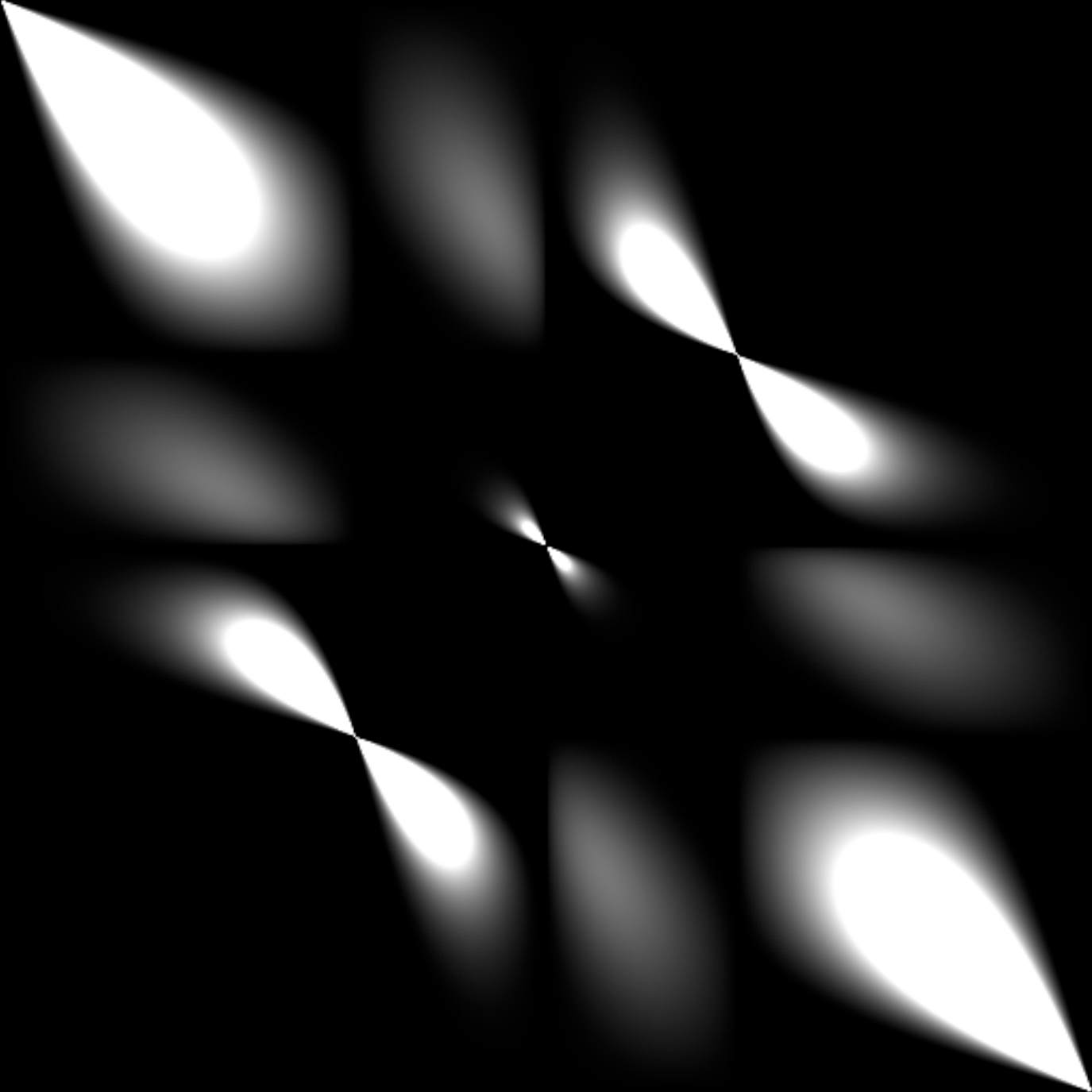}
     \includegraphics[width=3.5cm]{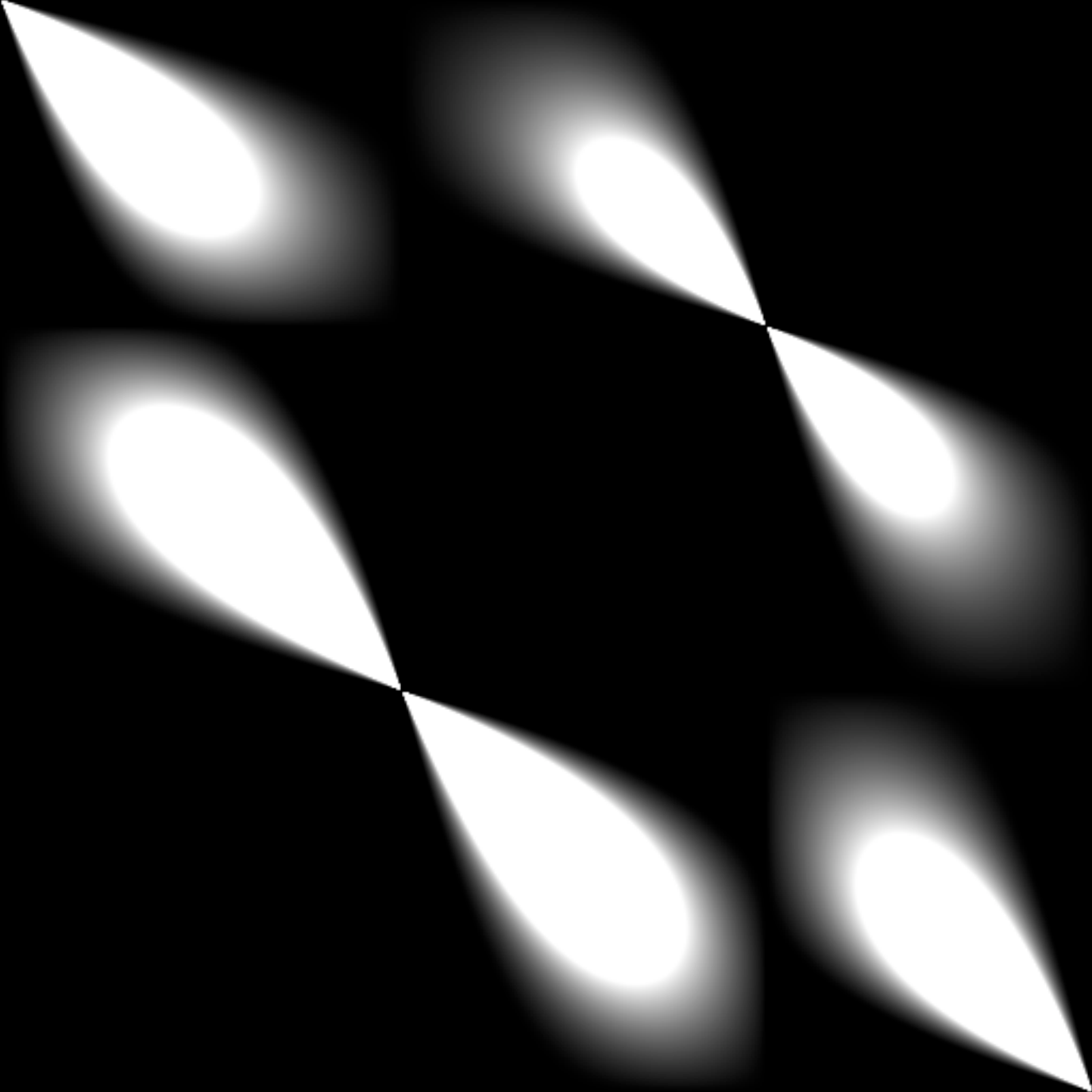} 
     \caption{Bayesian prior $P{(s_\inn,s_\out)}$ as a function of
       $s_\inn$ (horizontal axis) and $s_\out$ (vertical axis), where
       we have assumed isotropic source trajectories and uniform
       distributions for $\tE$ and event rate. Higher values of $P$
       appear in white (linear scale). From left to right,
       the caustic configurations are: (a) intermediate 
       with $d=1.1$ and $q=0.1$; (b) wide+central and (c) wide+secondary
       caustic, both for $d=2$ and $q=0.1$; (d) close+central and
       (e) close+secondary caustic, both for $d=0.5$ and $q=0.1$.}
     \label{fig:densities}
   \end{center}
 \end{figure*}

 As expected, the Jacobian $J$ is a function of
 the two parameters $(s_\inn,s_\out)$, while the
 bijection between the two set of parameters was possible by involving
 $(t_\inn,t_\out)$. However, $J$ is not yet the
 Bayesian prior $P{(s_\inn,s_\out)}$ we seek. We have
 yet to consider two aspects. Firstly, the parameters
 $(\uo,\alpha,\tE,\to)$ are themselves affected by prior
 probability distributions; this is discussed at the end of this
 section and is the topic of \Sec{sec:expriors}.
 Secondly, caustic crossing points
 are either entries or exits since the trajectory is orientated from
 $t_\inn$ to $t_\out$ ($t_\out \geq t_\inn$), which is not
 accounted for in \Eq{eq:Jacobfinal}. 
 To solve this second issue, we calculate the outward normal vector to the
 caustics at point $\zeta$, 
 \begin{equation}
   N_c = i\,\frac{d\zeta}{d\phi} \left/ \: \left|\frac{d\zeta}{d\phi}\right|\right.
 \label{eq:Nc}
 \end{equation}
 as well as the normalised and orientated trajectory vector
 \begin{equation}
   N_t =\frac{\zeta_\out-\zeta_\inn}{\left|\zeta_\out-\zeta_\inn\right|} \,,
   \label{eq:Nt}
 \end{equation}
 and check whether 
 $N_{c,\inn} \cdot N_{t,\inn} < 0$ (inward motion at $\zeta_\inn$) 
 and $N_{c,\out} \cdot N_{t,\out} > 0$ (outward motion at
 $\zeta_\out$). 
 If these conditions are fulfilled, we write $P{(s_\inn,s_\out)}=J$, 
 and $0$ otherwise. 

 Defined in this way, $P{(s_\inn,s_\out)}$ is thus
 the prior on $(s_\inn,s_\out)$ that we seek, in the special case
 of isotropic source trajectories (uniform distributions for $\uo$ and
 $\alpha$), uniform microlensing events rate ($\to$ is a random
 number), and uniform Einstein time $\tE \geq 0$. 
 In \Fig{fig:densities}, we have plotted $P{(s_\inn,s_\out)}$ for
 various $(d,q)$ configurations as a
 function of $s_\inn$ (horizontal axis) and $s_\out$ (vertical axis),
 higher values of $P$ appearing in white (linear scale).
 From left to right, these configurations are: (a) intermediate with 
 $d=1.1$ and $ q=0.1$; (b) wide+central and (c) wide+secondary, both
 configurations for $d=2$ and $q=0.1$; (d) close+central and (e)
 close+secondary caustic, both for
 $d=0.5$ and $q=0.1$. One can compare the intermediate case plot with 
 \Fig{fig:trajshoot}. In \Sec{sec:expriors}, we investigate how
 assuming different priors on the Einstein time $\tE$ affect the prior
 on $(s_\inn,s_\out)$. 
 
 \subsection{Extended sources} \label{sec:finitesrce}

 When the source approaches the caustic curves (at typically less than
 three projected source radii), one needs to take into account extended
 source effects in the modelling. As for $t_\inn$ and $t_\out$, 
 it is usually possible to extract from the light curve a new
 parameter that can be used instead of the source radius.

 It is well known that when the source crosses a straight
 line caustic (which is in many cases a good approximation of a real
 caustic), one can easily infer the duration of the crossing
 from the shape of the caustic crossing feature itself 
 \citep{Cassan69letter,Albrow1999,SchneiderWagoner1987}. Here, we
 define this duration as the time for the source to cross
 the caustic line by its full radius (\ie, from centre to limb), so that 
 $\Delta t_{\rm cc} = \rhoS/v_{\perp}$. In this definition,  
 $\rhoS$ is the source radius in Einstein ring radius units,
 $v_{\perp}$ is the component of the source velocity perpendicular to
 the caustic, and the subscript ``cc'' refers to either the caustic
 entry (``in'') or exit (``out''). For a given absolute velocity
 $1/\tE$, the source will take longer to cross the caustic if the
 trajectory makes a tangential angle with it. 
 More precisely, the normal velocity is
 proportional to the cosine of the angle between the trajectory and
 the caustic normal
 $v_{\perp} = |N_{c,{\rm cc}} \cdot N_{t,{\rm cc}}|/\tE$. 
 Inserting into this equation the expressions for 
 $\tE$, $N_{c,{\rm cc}}$, and $N_{t,{\rm cc}}$ (Eqs.~\ref{eq:tE},
 \ref{eq:Nc}, and \ref{eq:Nt}, respectively), we can compute the
 source radius $\rhoS$ as a function of $\Delta t_{\rm cc}$
 \begin{equation}
   \rhoS = \frac
   {\left|(\zeta_\out-\zeta_\inn) \wedge 
       \frac{d\zeta_{\rm cc}}{d\phi_{\rm cc}}\right|}
   {\left(t_\out-t_\inn\right)\left|\frac{d\zeta_{\rm cc}}{d\phi_{\rm cc}}\right|} 
   \,\Delta t_{\rm cc} \,.
   \label{eq:rhoDt}
 \end{equation}

 This expression would be exact if the crossed caustic were a
 perfect and infinite straight line. 
 In reality, however, caustic curves always have a
 curvature, and sometimes the source partly crosses a cusp. 
 Nevertheless, there is no arguing that $\Delta t_{\rm cc}$ is more
 suitable than $\rhoS$ for parameterising the observed caustic 
 crossing, since its rough value can be estimated from the light curve,
 in contrast to $\rhoS$.
 In practice, we choose the caustic crossing that provides the most
 comprehensive data coverage and which ressembles most closely a
 straight line caustic crossing to extract the starting value when
 fitting $\Delta t_{\rm cc}$.

 \section{Markov Chain Monte Carlo fitting} \label{sec:MCMC}

 \subsection{Examples of prior probability distributions} \label{sec:expriors}
 
 For a given set of fitting parameters
 $(s_\inn,s_\out,t_\inn,t_\out,\Delta t_{\rm cc})$,
 the prior of the probed model is given by
 \begin{equation} 
   P({\rm model})=
   P{(s_\inn,s_\out)} \, P(t_\inn, t_\out, \Delta t_{\rm cc}) \,.
   \label{eq:priortintout}
 \end{equation}
 The prior $P(s_\inn,s_\out)$ is computed as explained in
 \Sec{sec:analytic}, and may include priors that have been defined using
 the other parameters $\uo$, $\alpha$, $\tE$, $\to$, or $\rhoS$ by
 properly weighting $P(s_\inn,s_\out)$. Given Eqs.~(\ref{eq:tE}) and
 (\ref{eq:to}), a prior $P(t_\inn, t_\out)$ is equivalent 
 to a prior $P(\to, \tE)$ with a corresponding change in the prior 
 $P(s_\inn,s_\out)$. We now discuss different priors for
 the various parameters that could realistically be used in the Bayesian
 analysis. In \Fig{fig:densities} for example, we illustrate the case of isotropic
 trajectories, which corresponds to uniform priors for the parameters
 $\uo$ and $\alpha$. This choice is justifiable, since the direction of the
 binary lens axis is random. 

 \begin{figure}[!htbp]
   \begin{center}
     \includegraphics[width=5cm]{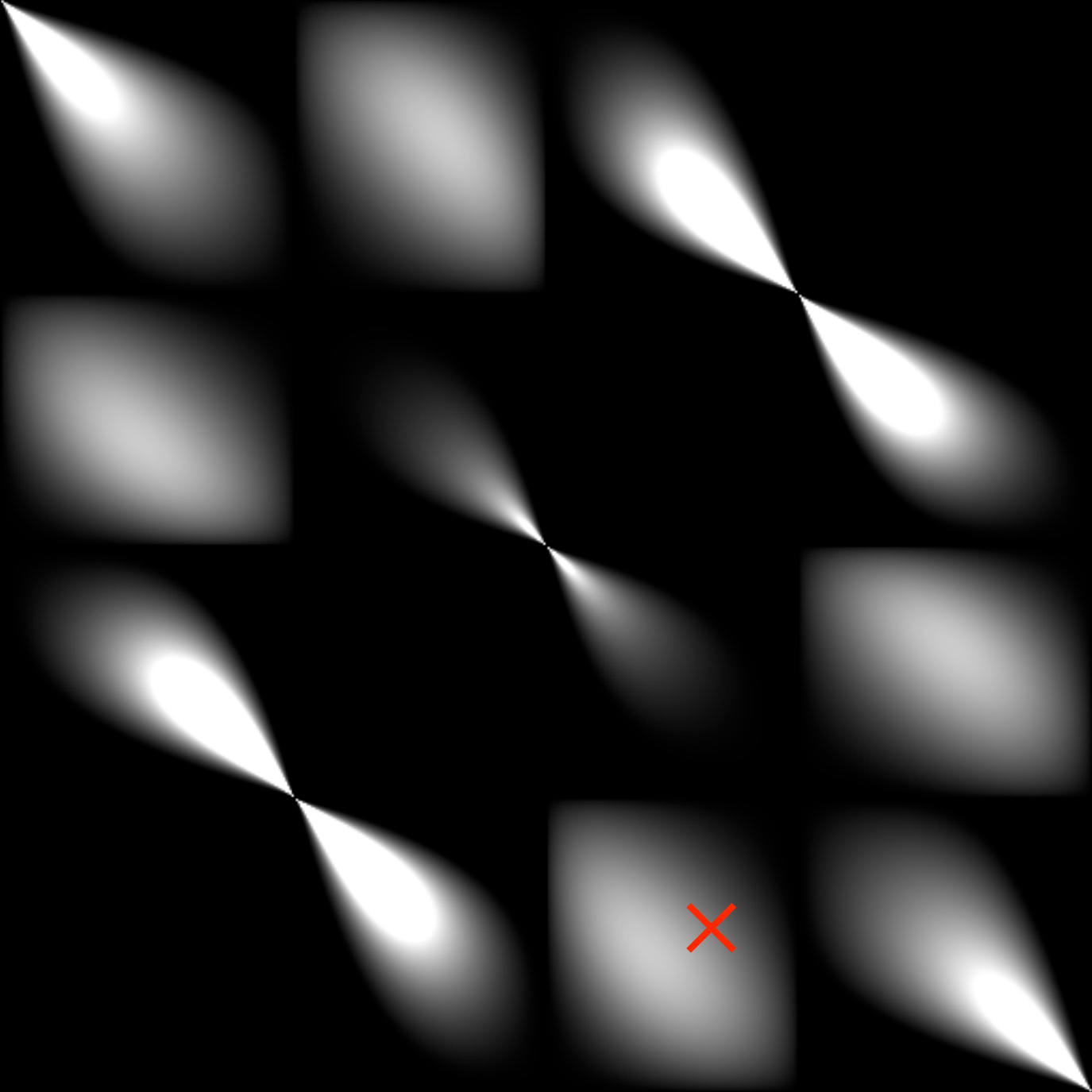} 
     \caption{
       Prior $P(s_\inn,s_\out)$ for the solution configuration of
       caustic crossing event OGLE~2002-BLG-069 (close+central
       caustic), with an underlying (uninformative) uniform prior in $\log\tE$.
       The red cross shows the location of the found caustic
       crossing, $(1.3,0.3)$, which falls in a region of relative high
       probability. }
     \label{fig:ob69}
   \end{center}
 \end{figure}

 The first class of priors that we can use are uninformative priors. Since
 the prior expresses information about the values of parameters before
 any data has been taken, we know that parameters such as
 $\to$, $\tE$, or $\Delta t_{\rm cc}$ must have uninformative priors,
 because we can only estimate their values by examining the light curve. 
 Although it is natural to use uniform priors for $\to$, $\alpha$ or
 $\uo$, for strictly positive parameters such as 
 $\Delta t_{\rm cc}$ or $\tE$, it is more suitable and commonly decided to
 use an uninformative prior that is uniform in the logarithm of the parameter.

 We illustrate the use of an uninformative prior (uniform priors in
 $\log\tE$, in $\uo$, $\alpha$, and $\to$) by computing 
 $P(s_\inn,s_\out)$ for the solution configuration
 of the binary lens event OGLE~2002-BLG-069
 \citep{Kubas2005,Cassan69letter}.
 The configuration for that event was that of a source
 crossing the central caustic of a close binary lens with parameters
 $d=0.46$, $q=0.58$ and  $t_\out-t_\inn \simeq 14.5$~days.
 The resulting prior $P(s_{\mathrm{in}}, s_{\mathrm{out}})$ is
 plotted in \Fig{fig:ob69}, where the red cross shows the location of the caustic
 crossings at $s_\inn\simeq 1.3$, $s_\out\simeq 0.3$. This falls
 within a region of high probability, meaning that the corresponding
 $P(s_\inn,s_\out)$ prior would have been a
 reasonable choice for this event. 

 The second class of priors are those that we can derive using
 information known before the event is observed. In microlensing, 
 a convenient parameter on which such a prior can be placed is the
 Einstein time $\tE$. This parameter depends on the
 relative distances between the source, the lens, and the observer, the
 kinematics of both the lens and the source and the lens' mass
 function. Combining all these data can help us to determine
 which ranges of values of $\tE$ are more likely to be
 observed. For the event OGLE-2007-BLG-472 \citep{Kains2009}, no prior
 information was included on $\tE$ (or the prior was assumed
 to be uninformative), which cause the best-fit models to have
 unrealistically long $\tE$.

 The method presented here can
 indeed be extended to include informative priors on parameters
 other than $\tE$, such as the source flux distribution, the
 blending light due to the lens, the relative proper motion of the
 source and lens, or the source-radius caustic crossing-time, but this
 would require us to link the analysis to a Monte Carlo model of the
 Galaxy. Although our approach can be generalised to these possible
 extensions, they are beyond the scope of the present
 paper. Using $\tE$ also has the advantage that its statistical
 distribution is fairly well-constrained by observed single-lens light
 curves, since this parameter is common to single- and binary-lens
 events.
 
 \begin{figure}[!htbp]
   \begin{center}
     \includegraphics[width=9cm, bb= 33 0 760 443]{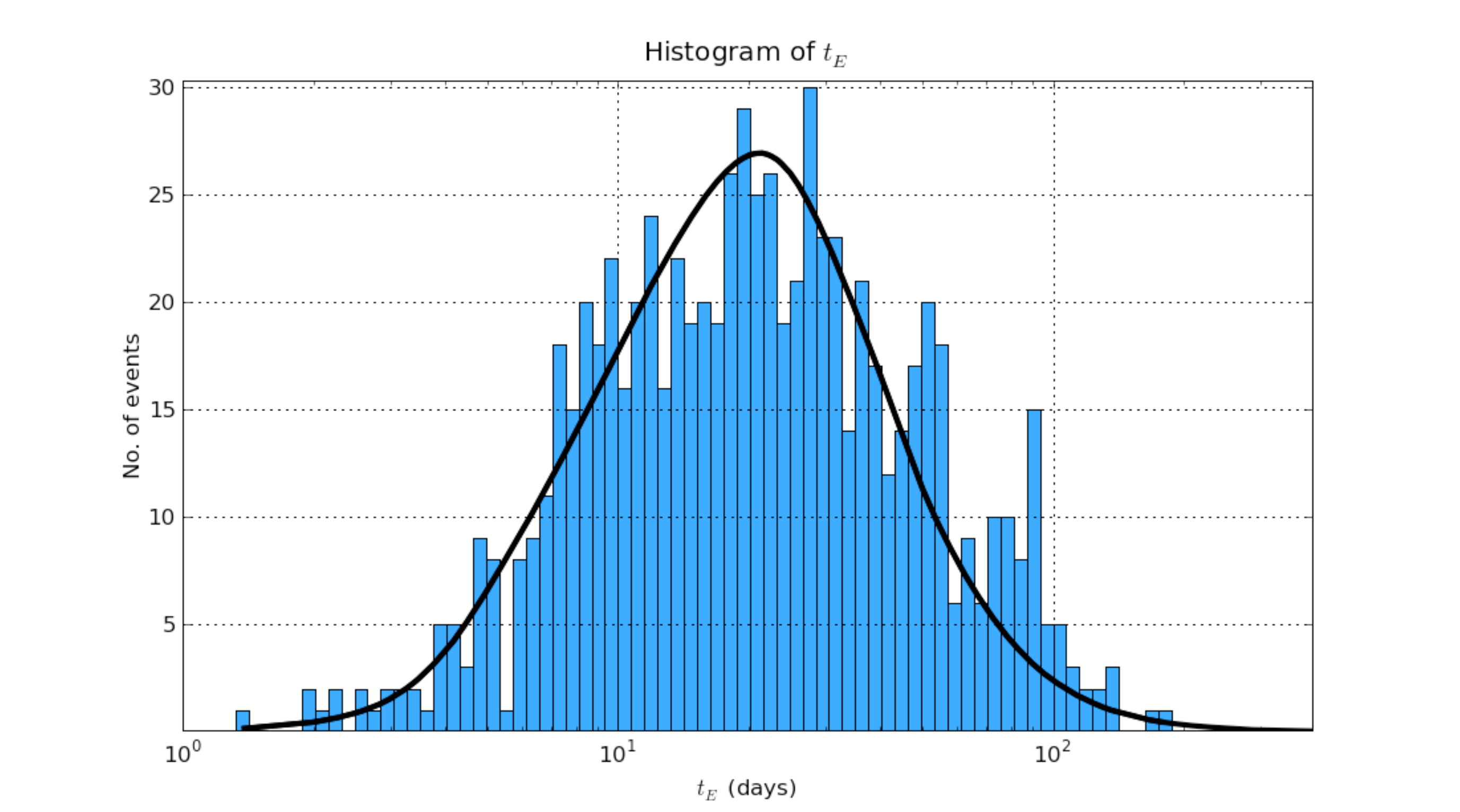} \\
     \vspace{0.3cm}
     \includegraphics[width=5cm]{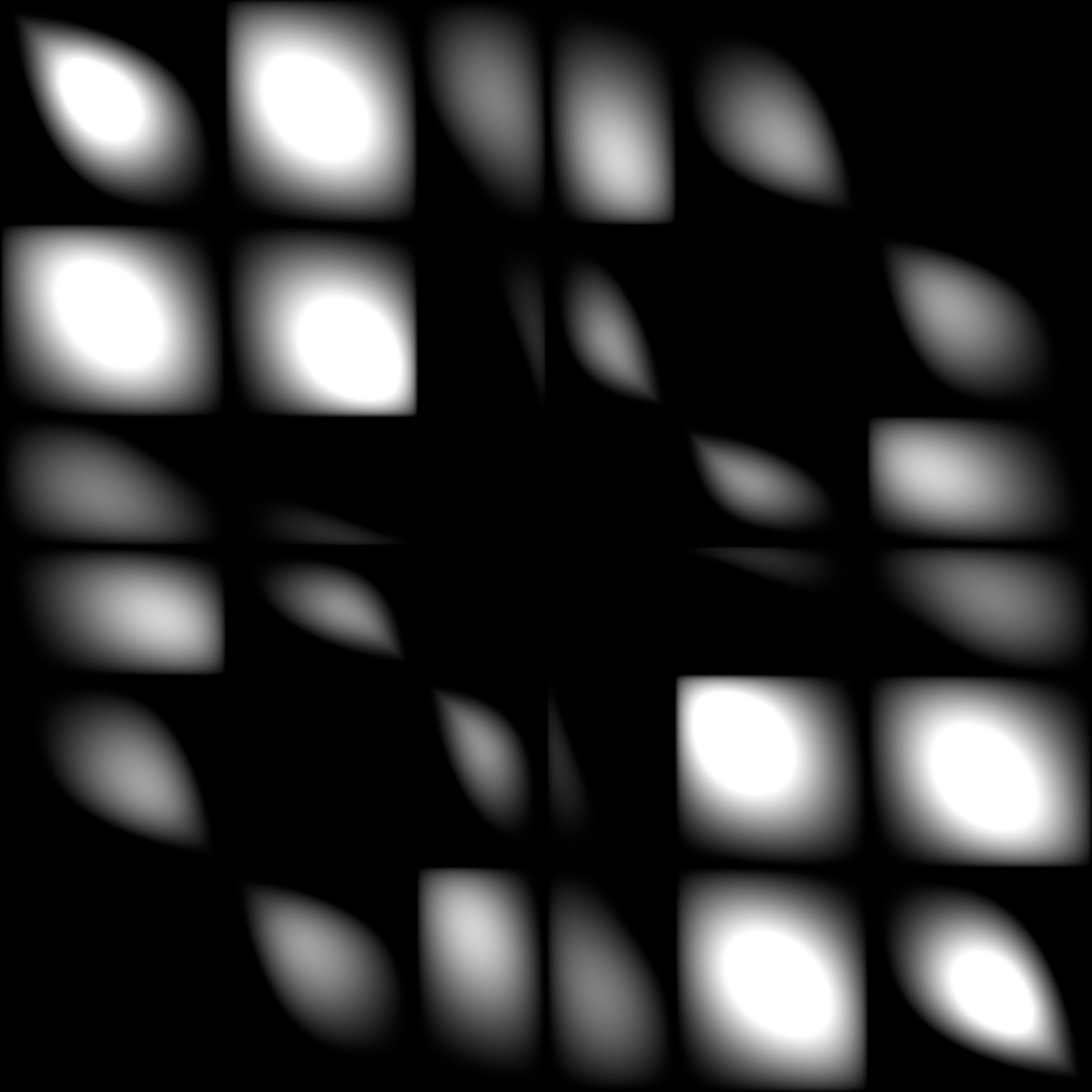} 
     \caption{
       In the top panel, the histogram (blue rectangles) shows the
       distribution of $\tE$ 
       found after fitting 788 single-lens microlensing events from
       OGLE 2006-2007 seasons. The solid black line shows the
       model prediction of \cite{WoodMao2005}, which is in good
       agreement with the data.
       The bottom panel displays the prior $P(s_\inn,s_\out)$ for the
       same intermediate caustic configuration as for
       \Fig{fig:densities} ($d=1.1$, $q=0.1$), but assuming an
       underlying prior for $\tE$ given by the above distribution. }
     \label{fig:WM}
   \end{center}
 \end{figure}
 
 Empirical distributions of $\tE$ can be obtained by
 modelling a large number of observed microlensing events. The top
 panel of \Fig{fig:WM} shows a histogram (blue rectangles) of
 $\tE$ values found by fitting 788 single-lens microlensing
 events from the 2006-2007 OGLE seasons (including blending). As
 expected, the distribution is far from uniform but instead appears
 roughly log-normal with a peak close to $\log\tE \simeq 1.32$
 and $\sigma_{\log\tE} \simeq 0.4$. Theoretical
 distributions of $\tE$ can also be based on predictions
 obtained with a Galactic model, such as the distribution 
 advocated by \cite{WoodMao2005}. This is plotted as a solid black
 line on top of our histogram (\Fig{fig:WM}, top panel) and is seen
 to closely match the empirical distribution. 
 Nevertheless, the distribution of \cite{WoodMao2005} lacks both
 extremely long events (say $\tE > 300$~days) that can be interpreted
 as black hole lenses, and extremely short events 
 (say $\tE < 3$~days) that can be interpreted as
 evidence of a population of free floating planets. But selection effects
 cause these extreme events to be under-represented in the observed $\tE$
 distribution, as can be seen in \Fig{fig:WM}. For these exceptional cases,
 special treatment would be required, for example using a prior on $\tE$
 that is more generous to extreme values in an attempt to compensate
 for selection effects. For most of binary lens events, however, a mild
 discrimination against black hole or loose planet lenses seems
 appropriate.

 Using the \cite{WoodMao2005} distribution as a prior, we compute and
 plot (\Fig{fig:WM}, bottom panel) the corresponding distribution
 $P(s_\inn,s_\out)$ by assuming $(t_\out-t_\inn) = 20$~days, $d=1.1$, and
 $q=0.1$ (the same intermediate configuration as \Fig{fig:densities}).
 Figure~\ref{fig:WM} (bottom panel) shows that with this prior, 
 cusp-crossing trajectories are far less
 likely to happen. For a trajectory near the cusps, this is because
 the source has only a short distance to travel between the entry
 and exit, while $(t_\out-t_\inn)$ is constant, meaning that the
 source's motion has to be very slow, leading to large 
 values of $\tE$, which are now ruled out
 by the prior\footnote{
   More precisely, when $\tE\rightarrow\infty$, 
   \cite{WoodMao2005} $\tE$ distribution behaves like $1/\tE^3 \sim
   |\zeta_\out-\zeta_\inn|^{3}$, and since $J\sim
   1/|\zeta_\out-\zeta_\inn|^2$, the net result is that near cusps,
   $J\sim|\zeta_\out-\zeta_\inn|\rightarrow 0$.}.
 This effect can be seen directly in the plot of
 $P(s_\inn,s_\out)$, where strong ``wing'' features at the cusps
 disappear, and other features appear (compare with \Fig{fig:densities}).

 \subsection{Posterior probability distributions: MCMC fitting}
 
 In practice, these and other statistics related to the posterior 
 parameter distribution can be
 evaluated efficiently using a Markov chain Monte Carlo
 to evaluate the probability-weighted integrals in Bayes' theorem. 
 A random walk in the parameter space is undertaken by
 taking random steps drawn from a distribution of the parameters $\theta$. 
 Each proposed step is accepted or rejected based on
 the probability of the new point relative to the old one exceeding
 some threshold, which is adjusted to maintain the acceptance rate
 above roughly $20$-$30$\%.
 The resulting chain locates and wanders around a local minimum,
 sampling the parameters with a weight proportional to
 the posterior probability.

 For a maximum likelihood analysis, the relative probability
 used to accept or reject new steps is $\exp{\left\{-\Delta\chi^2/2\right\}}$
 alone, where $\Delta\chi^2$ is the $\chi^2$ difference between the new and old 
 points; in a full Bayesian analysis, we multiply this exponential factor by the
 ratio of new to old values of the prior $P({\rm model})$, following 
 \Eq{eq:priortintout}.
 The posterior probability that the parameters $\theta$ lie in a defined 
 region $\Theta$ is then
 \begin{equation}
   P(\theta \in \Theta) = 
   \int_\Theta P(\theta|D)\, \d \theta  \,.
 \end{equation}
 The expected value of any function of parameters, $g(\theta)$, is
 \begin{equation}
   \left< g \right> \equiv \int g(\theta)\,
   P(\theta|D)\, \d \theta \,,
 \end{equation}
 and the variance about that expected value is
 \begin{equation}
   {\rm Var} \left[ g(\theta) \right] \equiv \int 
   \left( g(\theta) - \left< g \right> \right)^2\,
   P(\theta|D)\, \d \theta \,.
 \end{equation}
 In a similar way, confidence intervals, parameter covariances, and confidence 
 intervals can all be evaluated easily in the usual manner given the
 posterior probability distribution found with the MCMC algorithm,
 providing us with a complete statistical picture of the parameter
 space that we explore.

 \section{Conclusion}
 
 We have investigated plausible priors for Bayesian analysis of
 caustic-crossing microlensing light curves, based on an alternative
 parameterisation introduced by \cite{Causfix}. We have developed
 a mathematical formulation that allows us to compute
 analytically Bayesian priors for these parameters, given the knowledge
 we have about the physical quantities on which they depend. A number of
 relevant priors that may be used in a Bayesian, Markov chain Monte Carlo
 implementation of the given equations have been explored.

 In the context of the rapid development of a new generation of networks of
 classical and robotic telescopes \citep[e.g.,][]{Tsapras2008rn}, as
 well as space-based observations such as with the ESA
 project satellite Euclid \citep{Euclid2010}, a current
 challenge facing the microlens planet search community is to fully
 automate the fitting of binary lens light curves in real time, after
 having detected an anomaly \citep[e.g.,][]{HorSnodTsa09}. 
 This would enable anomalies that are
 detected in the observed light curves to be characterised as quickly
 as possible and for us to ascertain whether the anomalous behaviour is caused by a
 planet-mass companion of the lens star. Identifying parameters that
 could be estimated automatically by analysing the light curve (e.g., a
 magnification jump due to a caustic crossing) is already a step forward
 in accelerating the fitting codes by exploring a far more tighter
 parameter space. This was the motivation of \cite{Causfix} in defining
 a new set of parameters. In this work, we have added the possibility
 of including Bayesian priors in the analysis, which would avoid the
 need to explore combinations of parameters that are unlikely to
 happen.

 %
 
 \bibliographystyle{aa}
 \bibliography{13755}

\begin{thebibliography}{20}
\expandafter\ifx\csname natexlab\endcsname\relax\def\natexlab#1{#1}\fi

\bibitem[{{Albrow} {et~al.}(1999){Albrow}, {Beaulieu}, {Caldwell}, {Depoy},
  {Dominik}, {Gaudi}, {Gould}, {Greenhill}, {Hill}, {Kane}, {Martin},
  {Menzies}, {Naber}, {Pogge}, {Pollard}, {Sackett}, {Sahu}, {Vermaak},
  {Watson}, {Williams}, \& {The PLANET Collaboration}}]{Albrow1999}
{Albrow}, M.~D., {Beaulieu}, J.-P., {Caldwell}, J.~A.~R., {et~al.} 1999, \apj,
  522, 1022

\bibitem[{{Beaulieu} {et~al.}(2010){Beaulieu}, {Bennett}, {Batista}, {Cassan},
  {Kubas}, {Fouque}, {Kerrins}, {Mao}, {Miralda-Escude}, {Wambsganss}, {Gaudi},
  {Gould}, \& {Dong}}]{Euclid2010}
{Beaulieu}, J.~P., {Bennett}, D.~P., {Batista}, V., {et~al.} 2010,
  ArXiv/1001.3349

\bibitem[{{Beaulieu} {et~al.}(2006){Beaulieu}, {Bennett}, {Fouqu{\'e}},
  {Williams}, {Dominik}, {Jorgensen}, {Kubas}, {Cassan}, {Coutures},
  {Greenhill}, {Hill}, {Menzies}, {Sackett}, {Albrow}, {Brillant}, {Caldwell},
  {Calitz}, {Cook}, {Corrales}, {Desort}, {Dieters}, {Dominis}, {Donatowicz},
  {Hoffman}, {Kane}, {Marquette}, {Martin}, {Meintjes}, {Pollard}, {Sahu},
  {Vinter}, {Wambsganss}, {Woller}, {Horne}, {Steele}, {Bramich}, {Burgdorf},
  {Snodgrass}, {Bode}, {Udalski}, {Szyma{\'n}ski}, {Kubiak}, {Wi{\c e}ckowski},
  {Pietrzy{\'n}ski}, {Soszy{\'n}ski}, {Szewczyk}, {Wyrzykowski},
  {Paczy{\'n}ski}, {Abe}, {Bond}, {Britton}, {Gilmore}, {Hearnshaw}, {Itow},
  {Kamiya}, {Kilmartin}, {Korpela}, {Masuda}, {Matsubara}, {Motomura},
  {Muraki}, {Nakamura}, {Okada}, {Ohnishi}, {Rattenbury}, {Sako}, {Sato},
  {Sasaki}, {Sekiguchi}, {Sullivan}, {Tristram}, {Yock}, \&
  {Yoshioka}}]{OGLE05390Lb}
{Beaulieu}, J.-P., {Bennett}, D.~P., {Fouqu{\'e}}, P., {et~al.} 2006, \nat,
  439, 437

\bibitem[{{Bond} {et~al.}(2001){Bond}, {Abe}, {Dodd}, {Hearnshaw}, {Honda},
  {Jugaku}, {Kilmartin}, {Marles}, {Masuda}, {Matsubara}, {Muraki}, {Nakamura},
  {Nankivell}, {Noda}, {Noguchi}, {Ohnishi}, {Rattenbury}, {Reid}, {Saito},
  {Sato}, {Sekiguchi}, {Skuljan}, {Sullivan}, {Sumi}, {Takeuti}, {Watase},
  {Wilkinson}, {Yamada}, {Yanagisawa}, \& {Yock}}]{Ref-MOA}
{Bond}, I.~A., {Abe}, F., {Dodd}, R.~J., {et~al.} 2001, \mnras, 327, 868

\bibitem[{{Cassan}(2008)}]{Causfix}
{Cassan}, A. 2008, \aap, 491, 587

\bibitem[{Cassan {et~al.}(2004)Cassan, Beaulieu, Brillant, Coutures, Dominik,
  Donatowicz, Jrgensen, Kubas, Albrow, Caldwell, P., Greenhill, Hill, Horne,
  Kane, Martin, Menzies, Pollard, Sahu, Vinter, Wambsganss, Watson, Williams,
  Fendt, Hauschildt, Heinmueller, Marquette, \& Thurl}]{Cassan69letter}
Cassan, A., Beaulieu, J.~P., Brillant, S., {et~al.} 2004, \aap, 419, L1

\bibitem[{{Einstein}(1936)}]{Einstein1936}
{Einstein}, A. 1936, Science, 84, 506

\bibitem[{{Gould} \& {Loeb}(1992)}]{GouldLoeb1992}
{Gould}, A. \& {Loeb}, A. 1992, \apj, 396, 104

\bibitem[{{Horne} {et~al.}(2009){Horne}, {Snodgrass}, \&
  {Tsapras}}]{HorSnodTsa09}
{Horne}, K., {Snodgrass}, C., \& {Tsapras}, Y. 2009, \mnras, 396, 2087

\bibitem[{{Kains} {et~al.}(2009){Kains}, {Cassan}, {Horne}, {Albrow},
  {Dieters}, {Fouqu{\'e}}, {Greenhill}, {Udalski}, {Zub}, {Bennett}, {Dominik},
  {Donatowicz}, {Kubas}, {Tsapras}, {Anguita}, {Batista}, {Beaulieu},
  {Brillant}, {Bode}, {Bramich}, {Burgdorf}, {Caldwell}, {Cook}, {Coutures},
  {Dominis Prester}, {J{\o}rgensen}, {Kane}, {Marquette}, {Martin}, {Menzies},
  {Pollard}, {Rattenbury}, {Sahu}, {Snodgrass}, {Steele}, {Vinter},
  {Wambsganss}, {Williams}, {Kubiak}, {Pietrzy{\'n}ski}, {Soszy{\'n}ski},
  {Szewczyk}, {Szyma{\'n}ski}, {Ulaczyk}, \& {Wyrzykowski}}]{Kains2009}
{Kains}, N., {Cassan}, A., {Horne}, K., {et~al.} 2009, \mnras, 395, 787

\bibitem[{{Kubas} {et~al.}(2005){Kubas}, {Cassan}, {Beaulieu}, {Coutures},
  {Dominik}, {Albrow}, {Brillant}, {Caldwell}, {Dominis}, {Donatowicz},
  {Fendt}, {Fouqu{\'e}}, {J{\o}rgensen}, {Greenhill}, {Hill}, {Heinm{\"u}ller},
  {Horne}, {Kane}, {Marquette}, {Martin}, {Menzies}, {Pollard}, {Sahu},
  {Vinter}, {Wambsganss}, {Watson}, {Williams}, \& {Thurl}}]{Kubas2005}
{Kubas}, D., {Cassan}, A., {Beaulieu}, J.~P., {et~al.} 2005, \aap, 435, 941

\bibitem[{{Kubas} {et~al.}(2008){Kubas}, {Cassan}, {Dominik}, {Bennett},
  {Wambsganss}, {Brillant}, {Beaulieu}, {Albrow}, {Batista}, {Bode}, {Bramich},
  {Burgdorf}, {Caldwell}, {Calitz}, {Cook}, {Coutures}, {Dieters}, {Dominis
  Prester}, {Donatowicz}, {Fouqu{\'e}}, {Greenhill}, {Hill}, {Hoffman},
  {Horne}, {J{\o}rgensen}, {Kains}, {Kane}, {Marquette}, {Martin}, {Meintjes},
  {Menzies}, {Pollard}, {Sahu}, {Snodgrass}, {Steele}, {Tsapras}, {Vinter},
  {Williams}, {Woller}, {Zub}, \& {The PLANET/Robonet
  Collaboration}}]{Jovi2008}
{Kubas}, D., {Cassan}, A., {Dominik}, M., {et~al.} 2008, \aap, 483, 317

\bibitem[{{Mao} \& {Paczynski}(1991)}]{MaoPaczynski1991}
{Mao}, S. \& {Paczynski}, B. 1991, \apjl, 374, L37

\bibitem[{{Paczynski}(1986)}]{Paczynski1986}
{Paczynski}, B. 1986, \apj, 304, 1

\bibitem[{Schneider \& Wagoner(1987)}]{SchneiderWagoner1987}
Schneider, P. \& Wagoner, R.~V. 1987, \apj, 314, 154

\bibitem[{{Trotta}(2008)}]{TrottaMCMC2008}
{Trotta}, R. 2008, Contemporary Physics, 49, 71

\bibitem[{{Tsapras} {et~al.}(2009){Tsapras}, {Street}, {Horne}, {Snodgrass},
  {Dominik}, {Allan}, {Steele}, {Bramich}, {Saunders}, {Rattenbury}, {Mottram},
  {Fraser}, {Clay}, {Burgdorf}, {Bode}, {Lister}, {Hawkins}, {Beaulieu},
  {Fouqu{\'e}}, {Albrow}, {Menzies}, {Cassan}, \&
  {Dominis-Prester}}]{Tsapras2008rn}
{Tsapras}, Y., {Street}, R., {Horne}, K., {et~al.} 2009, AN, 330, 4

\bibitem[{{Udalski}(2003)}]{Ref-OGLE}
{Udalski}, A. 2003, Acta Astronomica, 53, 291

\bibitem[{Witt(1990)}]{Witt1990}
Witt, H.~J. 1990, \aap, 236, 311

\bibitem[{{Wood} \& {Mao}(2005)}]{WoodMao2005}
{Wood}, A. \& {Mao}, S. 2005, \mnras, 362, 945

\end{thebibliography}
\end{document}